\newcommand{\be}{\begin{equation}}
\newcommand{\ee}{\end{equation}}
\newcommand{\bea}{\begin{eqnarray}}
\newcommand{\eea}{\end{eqnarray}}
\newcommand{\ba}[1]{\begin{array}{#1}}
\newcommand{\ea}{\end{array}}

\newcommand{\diracslash}[1]{#1\llap{/\kern2pt}}

\documentclass[aps,amsmath,amssymb,twocolumn,floatfix,nobalancelastpage,footinbib,eqsecnum,pra,showpacs]{revtex4-1}
\usepackage{epsfig}
\usepackage{times}
\usepackage{mathrsfs}
\usepackage{graphicx}
\usepackage{epsfig}
\usepackage{dcolumn}
\usepackage{color}
\usepackage{bm}
\usepackage[caption=false]{subfig}
\usepackage[margin=1in]{geometry}

\begin{document}
\title{Optical Feshbach resonances through a molecular dark state:
Efficient manipulation of $p$-wave resonances in  fermionic  $^{171}$Yb atoms}
%\title{ Efficient $ p $-wave  Feshbach resonance  in   fermionic  $^{171}$Yb atoms induced by two lasers in strong-coupling regime}
\author{Subrata Saha$^1$, Arpita Rakshit$^1$, Debashree Chakraborty$^1$, Arpita Pal$^1$ and Bimalendu Deb$^{1,2}$}
\affiliation{$^1$Department of Materials Science, $^2$ Raman Centre for Atomic, Molecular and Optical Sciences, 
Indian Association
for the Cultivation of Science,
Jadavpur, Kolkata 700032, India.}
\begin{abstract}
In a recent experiment by Yamazaki {\it et al.} [Phys.Rev. A {\bf 87} 010704 (R) (2013) ], 
$p$-wave optical Feshbach resonance  in fermionic $^{171}$Yb atoms using  purely long-range molecular 
excited states 
has been demonstrated. We theoretically show that, if two purely 
long range excited states of $^{171}$Yb are coupled to the ground-state continuum of scattering states with two  
lasers, then it is possible to significantly suppress photoassociative atom loss by a dark resonance in the excited states.
 We present a general theoretical framework  for creating a dark state in electronically excited molecular potential
for the purpose of increasing the efficiency of an optical Feshbach resonance. This can be accomplished   
by properly adjusting the relative intensity, phase,
polarizations and frequency detunings of two lasers. 
We present selective numerical results on atom loss spectra, $p$-wave elastic and inelastic scattering 
cross sections of $^{171}$Yb atoms to illustrate the effects of the  molecular dark state on optical Feshbach resonance.  
\end{abstract}
\pacs{ 34.50.Cx, 34.50.Rk, 67.85.Lm}

\maketitle

\section{introduction}
Ability to control inter-particle interactions is important for exploring fundamental physics of many-particle 
systems in various interaction regimes. Towards this end, ultracold atomic gases offer unique opportunities since 
atom-atom interaction
at low energy can be manipulated with external fields. In recent times, a magnetic  Feshbach resonance (MFR) 
\cite{pra:1993:tiesinga,nature:1998:inouye,prl:1998:wieman} has been 
extensively used to tune $s$-wave scattering length $a_s$ of atoms over a wide range, facilitating the first demonstration of $s$-wave 
fermionic superfluidity in an atomic Fermi gas \cite{natue:2006:ketterle}.  $p$-wave MFR 
has been experimentally observed in spin polarized $^{40}$K \cite{prl:2003:bohn} and $^{6}$Li \cite{pra:2004:zhang,pra:2005} atomic gases, and theoretically analyzed  \cite{pra:2005:salomon, pra:2010:ueda}. MFR has been used to produced $p$-wave Feshbach molecules \cite{prl:2007:jin, prl:2007:stoof}.
Atom-atom interaction can also be altered by an 
optical Feshbach resonance (OFR) proposed by Fedichev {\it et al.} \cite{prl:1996:fedichev}.  
Tunability of $a_s$ by OFR  has been  experimentally demonstrated  \cite{prl:2000:fatemi,prl:2004:theis,pra:2005:thalhammer},
albeit for a limited range. Recently, Yamazaki {\it et al.} \cite{pra:2013:yamazaki} experimentally demonstrated $p$-wave OFR in fermionic $^{171}$Yb 
atoms following an earlier theoretical proposal by Goel {\it et al.} \cite{pra:2010:goel} .

In an OFR, a photoassociation (PA) laser is used to couple the scattering or free state of two $S$ (ground) atoms 
to a bound state 
in an excited molecular potential asymptotically connecting to one ground ($S$) and another excited ($P$) atom. 
The loss of atoms due to spontaneous emission from the excited bound state is a severe hindrance to efficient manipulation of 
atom-atom interactions by an optical method. In the dispersive regime, the magnitude of the free-bound detuning is larger
than both spontaneous and stimulated linewidths. In this regime, though the atom loss is mitigated,  
the change in elastic scattering  amplitude is small.
On the other hand, if the laser is tuned close to the free-bound transition frequency, there will be 
photoassociative formation of excited molecules which will eventually decay leading to drastic loss of atoms from the trap.
Till now 
it is found that OFR is not an efficient method for tuning $a_s$ as compared to MFR. It is therefore important to devise new methods to increase the efficiency of an OFR. Using an OFR one can manipulate not only $s$-wave but also higher partial wave interaction 
 \cite{prl:2009:deb, pra:2012:deb} between ultracold atoms. Apart from this, development of an efficient OFR method is primarily 
necessary to manipulate two-body interactions in nonmagnetic atoms to which an MFR is not applicable. 
The question 
then arises: Is there any way out to suppress the loss of atoms in order to 
coherently and all optically manipulate atom-atom interactions? To address this, we carry out a theoretical study  showing 
the manipulation of $p$-wave interaction in fermionic $^{171}$Yb atoms by two lasers in different coupling regimes. 
 
  \begin{figure}
            \includegraphics[height=1.5in,width=0.48\textwidth]{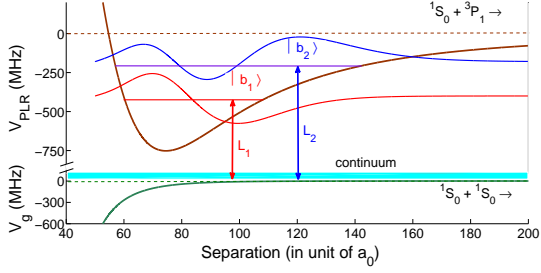}
            \caption{(Color online) A schematic diagram for generating coherence between two PLR bound states $\mid b_1 \rangle$ 
            and  $\mid b_2 \rangle$   of $^{171}$Yb$_2$ by two OFR lasers $L_1$ and $L_2$.  The ground-state 
            potential $V_g$ (in unit of MHz) and PLR potential $V_{{\rm PLR}}$ (in unit of MHz) are plotted 
            against the internuclear separation in unit of Bohr radius $a_0$. Asymptotically, $V_g$
            corresponds to two separated atoms in electric $^1S_0 + ^1S_0$ states while $V_{{\rm PLR}}$ connect to 
            two separated atoms in electronic $^1S_0 + ^3P_1$ states. $\mid b_1 \rangle$ and $\mid b_2 \rangle$ have 
            vibrational quantum number $\nu_1 = 1$ and $\nu_2 = 2$, respectively; both have the same 
            rotational quantum number $T_e $ equal to either 1 or 3.}      
            \label{fig:ccoh}
             \end{figure}

   \begin{figure}
           \includegraphics[height=1.5in,width=0.48\textwidth]{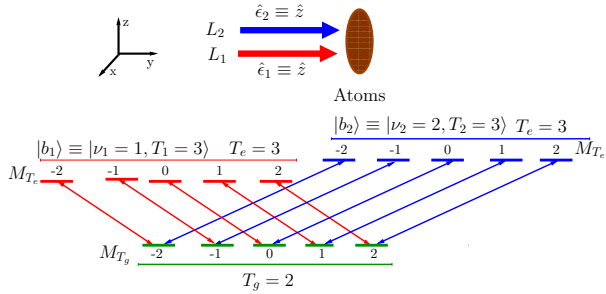}
           \caption{(Color online) A schematic diagram showing how two lasers couple different molecular magnetic sub-levels of ground and excited states of 
           $^{171}$Yb atoms.  Both lasers are linearly polarized along the $z$-axis. A pair of photons - one from L$_1$ (red) and the other 
           from L$_2$ (blue) laser, couple a particular ground state magnetic sub-level $M_{T_g}$ with 
           two excited states having same angular momenta $T_e = 3$ and  $M_{T_e} = M_{T_g}$, where $M_{T_g}$ can take any of  the 5 values  
           from -2 to 2. Since the angular parts of the amplitudes for each such pair of transitions are the same, the magnitudes of the two 
           transition amplitudes can be made equal by suitably adjusting the relative intensity of the two lasers. The phase of the two transition 
           amplitudes is opposite due to the opposite vibrational parity of the two excited states.  }      
           \label{figmt}
            \end{figure}           
Here we show that it is possible to make an OFR significantly efficient by substantially suppressing  atom loss   
by the method of dark state resonance \cite{nuovocimento:1976:alzetta,propt:1996:arimondo} in molecular 
excited states. Usually, a dark 
state resonance refers to the formation of a coherent superposition   
of two ground-state sub-levels of an atom or a molecule by two lasers.            
When a dark state is formed in ground-state sub-levels, an atom or a molecule  
can not effectively absorb a photon to reach to an excited state, and therefore no fluorescence 
light comes out.   
Dark resonance in ground-state sub-levels is well-known and plays an essential role in a number of 
 coherent phenomena such as coherent population trapping (CPT) \cite{book:atom-photon:1992:cohen-tan}, laser cooling 
 \cite{prl:1988:cohen-tan,josab:1989:cohen-tan}
 electromagnetically induced transparency (EIT) \cite{propt:1996:arimondo}, stimulated 
 Raman adiabatic passage (STIRAP) \cite{pra:1989:bergmann,aamop:2001:vitanov,rmp:1998:bergmann}, 
 slow light \cite{phystoday:1997:harris,nature:1999:hau} etc.. In contrast, 
 dark resonances in  atomic or molecular excited 
states have remained largely  unexplored,  because the excited states are in general too lossy. 
Now, with the accessibility of relatively long-lived excited  states of alkaline earth-like atoms 
via intercombination PA transitions \cite{pra:2005:julienne,prl:2008:enomoto1},  
it is possible to create a coherent superposition or a coherence in 
 ro-vibrational states of an excited molecule by two PA lasers. Though this laser-induced coherence 
has been discussed earlier in the 
 contexts of $d$-wave OFR \cite{pra:2012:deb}, vacuum-induced coherence \cite{pra:2012:rakshit} and rotational 
 quantum beats \cite{jpb:2014:rakshit}, here we give an exposition of the crucial role it can play in 
 manipulating an OFR. When two molecular ro-vibrational states belonging to the same 
 electronically excited potential have the same rotational  but different 
vibrational quantum numbers, it is possible to generate a superposition of  these two levels 
by applying two PA lasers. Under appropriate conditions, 
this superposed state can be protected against 
spontaneous emission leading to CPT in excited states \cite{pra:2012:rakshit}.

  Our purpose is to make use of  molecular dark resonance to control OFR. For illustration,            
 we investigate the manipulation of  $p$-wave scattering properties of  $^{171}$Yb atoms with 
two OFR lasers L$_1$ and L$_2$ tuned to PA transitions to two purely long-range (PLR) bound states $\mid b_1 \rangle$ and $\mid b_2 \rangle$, 
respectively; as shown in Fig. 1. $\mid b_1 \rangle$ and $\mid b_2 \rangle$ have been chosen to have the same 
rotational quantum number $T_1 = T_2 = T_e$ 
but different vibrational quantum numbers $\nu_1 = 1$  and $\nu_2 = 2$, respectively. 
$T$ may be chosen to be 1 or 3. 
By treating both lasers on an equal footing, 
we obtain results for any arbitrary optical-coupling regime. Our results show that 
the elastic scattering rate can exceed the inelastic rate by 5 orders of magnitude under the 
molecular dark resonance conditions in the strong-coupling regime. The atom loss can be almost completely eliminated by the use of the 
dark state. This leads to the huge enhancement in the efficiency of an all optical method 
for controlling atom-atom interactions. 
 
 The remainder of the paper is organized in the following way. In the next section we present our theory of two-laser OFR 
 emphasizing on the essential idea behind the creation and utilization of molecular dark resonance for suppression of 
 atom loss. In Sec.III, we apply this theoretical method to manipulate $p$-wave interactions of $^{171}$Yb atoms. We then present numerical 
 results and interpret them in Sec.IV. The paper is concluded in Sec.V.

\section{ Theory} 
\hspace{0.1in} We first give the general idea behind  molecular dark state-assisted OFR.
We then specialize our theoretical method for $p$-wave OFR in fermionic 
$^{171}$Yb atoms. A scheme of two-laser OFR in $^{171}$Yb has been  depicted in Fig. 1 which may also be 
used for a general scheme of two-laser OFR of ultracold atoms. Since photoassociative 
interaction between a nonzero partial-wave $(\ell \ne 0)$  scattering state of two electronically ground-state atoms and 
an excited molecular bound state is essentially anisotropic,  
the usual theoretical 
method for $s$-wave OFR given by Bohn and Julienne \cite{pra:1999:bohn} needs to be extended to 
include anisotropic effects. In 
our theoretical treatment, we take into account all the magnetic sub-levels of the ground and excited rotational levels. The 
two lasers are taken to be co-propagating  and linearly polarized along the $z$-axis.

Two  molecular bound states $\mid b_1 \rangle$ and  $\mid b_2 \rangle$
supported by an excited-state potential are coupled to the continuum of unperturbed ground scattering states  
$\mid E '\rangle$ with collision energy $E'$ by two PA lasers 
L$_1$ and L$_2$. 
This leads to the formation of energy-normalized dressed state \cite{pra:2012:deb}.
\begin{align} 
\mid \psi_E \rangle = A_{1E} \mid b_1 \rangle + A_{2E} \mid b_2 \rangle + \int d E' C_{E'}(E) \mid E' \rangle 
\label{dcont}
\end{align} 
where $A_{1E}$, $A_{2E}$ and $C_{E'}(E)$ are the dressed amplitudes given by 
\begin{align} 
A_{1E} = D_{1E}^{-1} e^{i\phi_{L_1}} \left [ \Lambda_{1E} + {\mathscr L}_1^{(2)} \right ]
\end{align} 
\begin{align} 
A_{2E} = D_{2E}^{-1} e^{i\phi_{L_2}} \left [ \Lambda_{2E} + {\mathscr L}_2^{(1)} \right ]
\end{align} 
\begin{align} 
C_{E'}(E) = \delta(E-E') + \int d E'  \frac{A_{1E}\Lambda_{1E'} + A_{2E}\Lambda_{2E'}}{E-E'} 
\end{align} 
where $\phi_{L_n}$ is the phase of L$_n$ laser. In the expression of $A_{nE}$, the first term   $D_{nE}^{-1} \Lambda_{nE}$ is the amplitude that depends on direct free-bound coupling $\Lambda_{nE}$ between the  bound state $\mid b_n\rangle$ and the continuum due to laser
L$_n$   while  the second term $D_{nE}^{-1} {\mathscr L}_n^{(n')}$ results from the cross-coupling between the bound states by the two lasers. Here the denominator $D_{n E} = E + \hbar \tilde{\delta}_n(E) + i \hbar G_n(E)$ with 
$\tilde{\delta}_n$ being the detuning
of the L$_n$ laser form the light-shifted $n^{\text{th}}$ bound state, 
\begin{align} 
\tilde{\delta}_n = \delta_n + \frac{1}{\hbar} ( E_{b_n}^{{\rm shift}} + E_{nn'}^{{\rm shift}})
\end{align} 
where $\delta_n=\Delta_n + E_{b_n}/ \hbar$ with $ \Delta_n = \omega_{L_n} - \omega_A  $ being the detuning of $n^{\text{th}}$ laser from the frequency of atomic transition 
$S \rightarrow P$, $E_{b_n}$ is the binding energy of $\mid b_n\rangle$ measured from the threshold of the 
excited-state potential and $E_{b_n}^{{\rm shift}} $ is the light-shift of the bound state. Here $G_n(E)= \Gamma_n(E) + \Gamma_{n n'}(E)$  with $\Gamma_n(E)$ being 
the stimulated line width of $n^{\text{th}}$ bound state due to L$_n$ laser,  $E_{n n'}^{{\rm shift}}$ and $\Gamma_{n n'}$ are the 
terms that depends on the terms  ${\mathscr L}_1^{(2)}$ and   ${\mathscr L}_2^{(1)}$. Note that 
$\Gamma_{n n'}$ can be negative, but $G_n \ge 0$.

Equation (1) is derived without taking spontaneous emission into account. 
Following Bohn and Julienne \cite{pra:1999:bohn}, 
spontaneous decay can be included into the problem  by introducing an `artificial' open channel in the ground state manifold.
Let the state of this artificial decay channel be denoted by $\mid E \rangle_{{\rm art}}$ and its interaction with an excited state by 
$V_{ {\rm art}}$.  The spontaneous  emission linewidth can be identified  with 
\begin{align}
\gamma_n = \frac{2 \pi}{\hbar} | \,  _{{\rm art}} \!\langle {\rm E } \mid V_{ {\rm art}} \mid b_n \rangle|^2
\end{align}
As a result, $G_n$ should be replaced by $G_n + \gamma_n$. The $T$-matrix element for inelastic process of 
transitions from the two correlated excited states  
to $\mid E \rangle_{{\rm art}}$ is  
\begin{align} 
\pi \, _{{\rm art}}\! \langle E \mid  
V_{{\rm art}} \mid \psi_E \rangle =  \pi \left ( {\mathcal V}_1 A_{1E} + {\mathcal V}_2 A_{2E} \right ) 
\end{align} 
where $ {\mathcal V}_n = \, _{{\rm art}}\! \langle E \mid V_{ {\rm art}} \mid b_n \rangle$. This gives the inelastic scattering 
cross section 
\begin{align} 
\sigma_{\rm{inel}} &=& \frac{4 \pi g_s}{ k^2 }\mid  \pi \left ( {\mathcal V}_1 A_{1E} + {\mathcal V}_2 A_{2E} \right )  \mid ^2
\end{align}
where $g_s = 1(2)$ for two distinguishable(indistinguishable) atoms. 
The atom loss rate is given by $K_{{\rm loss}} = \langle v_{rel} \sigma_{{\rm inel}} \rangle$ where $\langle \cdots \rangle$ implies 
thermal averaging over Maxwell-Boltzmann distribution of the relative velocity $v_{rel}=\hbar k/\mu $ with $k$ being the relative 
wave number. Clearly, $\sigma_{\rm{inel}}$ or $K_{{\rm loss}}$ will vanish if $ {\mathcal V}_1 A_{1E} = - {\mathcal V}_2 A_{2E}$, meaning  
\begin{align} 
\frac{A_{1E}}{A_{2E}} = - \frac{{\mathcal V}_2}{{\mathcal V}_{1}} = - \sqrt{\frac{\gamma_2}{\gamma_1}}
\end{align} 
is the condition for the onset of an excited  
molecular dark state that is protected against spontaneous emission. 
 This condition can be fulfilled by suitably 
adjusting the relative intensity and phase  between the two lasers,  and the two detuning parameters $\Delta_1$ and $\Delta_2$.

The elastic scattering amplitude can be obtained from asymptotic analysis of the dressed wavefunction
$ \psi_E ({\mathbf r}) = \langle {\mathbf r} \mid \psi_E \rangle$ where ${\mathbf r}$ stands for the relative position vector 
of the two atoms. This can be conveniently done by partial-wave expansion as done in detail in Ref. \cite{pra:2012:deb}. Scattering 
properties of low-lying 
partial wave $\ell = 0$ ($s$-wave) or  $\ell=1$ ($p$-wave) or $\ell=2$ ($d$-wave) can be optically manipulated by this two-laser OFR
method.   
Which partial-wave will be most influenced by this method depends on the rotational quantum numbers of the two excited bound states 
and the temperature of the atomic cloud. While ultracold temperatures in the Wigner threshold law-regime are most suitable for 
manipulating $s$-wave collisions, temperatures slightly higher than   Wigner threshold law-regime  or temperatures near a shape 
resonance are appropriate for controlling higher partial-wave collisional properties. As an illustration we analyze manipulation of 
$p$-wave scattering properties of $^{171}$Yb atoms with two-laser OFR in the next section.

\section{Two-laser $p$-wave OFR in $^{171}$Y\lowercase{b} atoms}

For the bound states $\mid b_1 \rangle $ and $\mid b_2 \rangle $, we choose purely long range (PRL) molecular 
states of $^{171}$Yb$_2$. These states are fundamentally different from usual molecular bound states on several 
counts. First, these states are formed due to an interplay between resonant dipole-dipole and spin-orbit  or hyperfine interactions  
in the excited atomic states. Second, 
their equilibrium position lies at a large separation well beyond the chemically interactive region 
of the  overlap between the electron clouds of the two atoms. 
Third, the constituent atoms retain most of their atomic characters.
Fourth, the potentials supporting such states are usually very shallow allowing only a few vibrational 
levels to exist. Predicted about 35 years ago  
\cite{jpb:1977:movr,prl:1978:stwalley}, these states have been recently experimentally observed in alkali-metal \cite{prl:1994:cline,jcp:1994:ratliff,pra:1996:stwalley,prl:1998:fioretti,pra:1998:wang,epjd:2000:comparat},
metastable helium \cite{prl:2003:cohen-tan} and fermionic ytterbium \cite{prl:2008:enomoto}  atoms with PA spectroscopy.  
These states will be  particularly useful for optical manipulation of
$p$- or higher partial-wave atom-atom interactions. One of the major obstacles to higher 
partial-wave OFR stems from the fact that the partial-wave centrifugal barrier is too high 
for low-energy scattering wavefunction to be appreciable in the short-range region. 
With PLR states being used for OFR, PA transitions need not take place 
inside the barrier, opening up new scope for higher partial-wave OFR. 
The fact that the photoassociative atom loss  mostly occurs at relatively short-range region 
makes PLR states a better choice for OFR in order to mitigate the loss.

In case of 
$^{171}$Yb atoms, PLR bound states are formed due to an interplay between 
resonant dipole-dipole and  hyperfine  interactions.
 The PLR potential  of $^{171}$Yb$_2$ of the state $^1S_0+^3P_1$ is obtained \cite{prl:2008:enomoto} by diagonalizing the adiabatic Hamiltonian 
  \begin{align}
  H_{{\rm adia}}=\frac{{\mathbf d}_1.{\mathbf d}_2-3d_{1z}d_{2z}}{R^3} + a({\mathbf i}_1.{\mathbf j}_1 + {\mathbf i}_2.{\mathbf j}_2)-\frac{C_6}{R^6}
  \label{diag}
\end{align}  
where  ${\mathbf{d}}_n$, ${d}_{nz}$, $\mathbf{{j}_n}$ and $\mathbf{{i}_n}$ denote the dipole moment of the atomic transition $^1S_{0} \rightarrow \, ^3\!P_1$, 
$z$-component of the dipole moment, electronic and nuclear spin angular momentum, respectively, of the $n^{\text{th}}$ (=1,2) atom. Here the parameters 
$a = 3957$ MHz,  $C_6 = 2810$ a.u. and the magnitude of the transition dipole moment $ d = 0.311$ a.u.. 
The axial projection $\Omega$ of
of the total electronic angular momentum ${\mathbf J} = {\mathbf j}_1 + {\mathbf j}_2$ is not a good quantum number. The total nuclear spin 
angular momentum is ${\mathbf I} = {\mathbf i}_1 + {\mathbf i}_2$. 
  The axial projection ($\Phi$) of the total angular momentum ${\mathbf F} = {\mathbf J} + {\mathbf I}$  is a good quantum number. 
 Further,  when we include the rotation of the internuclear axis described by the partial-wave $\ell$, 
 the  good quantum number are  ${\mathbf T} = {\mathbf F}
  + \vec{\ell} $ and its projection $M_T $ on the space-fixed axis.  
  Now,  effective  excited  potential is
  \begin{align}
 V_e(r)=V_{{\rm PLR}}+\frac{\hbar^2[T_e[T_e+1]+\langle F^2\rangle-2\varPhi^2}{2\mu r^2}
 \end{align}
 where $V_{{\rm PLR}}(r)$ is  the PLR potential obtained by diagonalizing  Eq.(\ref{diag}),  $\mu$ is the reduced mass 
 and $T_e$ represents the rotational quantum number of the excited molecular state. 
 The PLR potential that is accessible from $p$-wave ground-state scattering state via PA has 
 depth of about 750 MHz and equilibrium separation at 75 Bohr radius \cite{prl:2008:enomoto}. 
 The ground state potential is
 \begin{align}
 V_g=\frac{C_{12}}{r^{12}}-\frac{C_{6}}{r^{6}}-\frac{C_{8}}{r^{8}}+\frac{\ell(\ell+1)}{2\mu r^2}
 \end{align}
where $C_{12}=1.03409\times10^9 $ a.u, $C_{8}=1.93\times10^5 $ a.u and $C_{6}=1931.7 $ a.u 
\cite{pra:2010:goel,pra:2008:kitagawa}. Ground-state $^{171}$Yb has nuclear spin
$i=\frac{1}{2}$ with no electronic spin.  
The ground-state collision in $s ( p)$ partial-wave is characterized by the total nuclear spin
$I= 0 (1) $ due to the antisymmetry of total wave function. 
For molecular  transitions,  ground and excited states must have opposite parity. 
The selection rule is $\Delta T = 0, 1 $ with $ T = 0 \rightarrow T = 0$  being forbidden. 
For $\Phi_e \ne 0$ all $T_e$ are allowed. For $\Phi_e = 0$ state only  odd or even $T_e$ are allowed 
\cite{prl:2008:enomoto,pra:1996:Abraham,pra:2005:Tiesinga,Book1}. 
For $p$-wave we use  0$^-$ PLR state,  so only odd $T_e$ are allowed.

Let us consider two PLR excited bound states of $^{171}$Yb$_2$ represented by $ \mid b_{\nu} \rangle 
\equiv \mid T_e{M_T}_e\rangle_\nu $,
where $T_e$ is rotational quantum number, $M_{T_e}$ denotes its projection along the z-axis of the laboratory frame 
and $\nu(=1,2)$ is the vibrational quantum number. As shown in  Fig. 1, two OFR lasers  L$_1$ and L$_2$ are used to induce 
photo-associative dipole coupling between the  ground continuum state and 
the excited bound states  with vibrational quantum numbers $\nu =1$ and $\nu=2$, respectively. 
Both the bound states have the same 
rotational quantum numbers either $T_e =1$ or $T_e = 3$. Let the internal (rotational) state of the two ground-state atoms in 
molecular basis be denoted by $|T_g {M_T}_g\rangle $.

The dressed continuum of Eq. (\ref{dcont}) can be derived   
following the method given in the appendix-A of Ref. \cite{pra:2012:deb}. It can be most conveniently 
done using the expansion in terms of molecular angular momentum basis functions or spherical harmonics. However, instead of 
$\mid \ell, m_{\ell} \rangle $ basis for the ground state, we use $\mid T_g, M_{T_g} \rangle $ basis for the present context. 

The photoassociative loss of atoms is governed by the equation 
$ \dot{n} = K_{loss} n^2$ where $n$ is the atomic number density. Assuming a uniform number density $\bar{n}$  
the number of atoms remaining after the simultaneous action of both the lasers for a duration of $\tau$ 
is given by 
\begin{align} 
N_{\rm{remain}} = \frac{N_{0}}{ 1 + \bar{n} K_{{\rm loss}} \tau}
\end{align} 
where $N_0$ is the initial number of atoms. 
\subsection*{Elastic scattering cross section} 
For the geometry and polarizations chosen for the two laser beams as schematically shown in 
Fig. 2, it is clear that the optical transitions couple ground and excited magnetic sub-levels 
$M_{T_1} = M_{T_2} = M_{T_g} $. We take $T_1 = T_2 = T_e$.

The asymptotic form of $ \psi_E({\mathbf r})$ 
    is given by 
    \bea 
    \psi_E( {\mathbf r} \rightarrow \infty)&{\sim}&r^{-1} \sum_{M_{T_g}, M_{T_g}'} \psi_{T_g M_{T_g}}^{T_g M_{T_g}'}(r)Y_{T_g, M_{T_g}'}^*(\hat{k})\nonumber\\ &&\times\langle \hat{r} \mid T_g M_{T_g} \rangle
    \eea 
where 
\begin{widetext}
% \begin{align}
\bea
 \psi^{T_g M_{T_g}'}_{T_g M_{T_g}}(r) \langle \hat{r} \mid T_g M_{T_g} \rangle & = &  \sum_{\substack{m_l,m_I\\m_l+m_I=M_T}}
 {\mathcal C}^{T_g M_{T_g}}_{lm_lIm_I} \langle \hat{r}  |lm_l Im_I, T_g M_{T_g} \rangle \left [\sin\left(kr-\frac{l\pi}{2}-\eta \right)  -\pi F_{T_g M_{T_g}}^{T_{g} M_{T_g}'}(E,E')\right.\nonumber\\
 && \left.\times \exp \left \{ i\left(kr-\frac{l\pi}{2}-\eta\right) \right \}\right] 
% \end{align}
\eea
\end{widetext}
 where $\eta$ is the background (in the absence of lasers) phase shift and ${\mathcal C}^{T_g M_{T_g}}_{lm_lIm_I}$ is 
 Clebsch-Gordan coefficient.  Comparing this with Eq.(1) and using the expansion  
 $C_{E'} \equiv \sum_{M_{T_g},M_{T_g}'} C_{E', T_g M_{T_g}}^{T_g M_{T_g}'} Y_{T_g M_{T_g}'}^*(\hat{k})$ we can relate 
 $ C_{E',T_g M_{T_g}}^{T_g M_{T_g}'}\langle r \mid E' \rangle_g \equiv   \psi_{T_g M_{T_g}}^{T_g M_{T_g}'}(r)$.
  The ${\cal T}$ matrix element is given by 
 \begin{align}
 {{\cal T}}_{TM_T,T M_T'}=-
 \mathrm {e}^{i\eta} \sin\eta  \hspace*{0.1cm} \delta_{m_lm_l'}\delta_{m_Im_I'}
 +\pi F_{T_gM_{T_g}}^{T_g M_{T_g}'}\mathrm{e}^{2i\eta}.
 \end{align}
 Elastic scattering cross section is
 \begin{align}
  \sigma_{el}=\frac{4\pi g_s}{k^2}\sum_{M_T, M_{T_g}'}{| {{\cal T}}_{T_g M_{T_g},T_g M_{T_g}'}|}^2.
  \label{elas}
 \end{align}
 By using the expansion 
  $A_{\nu E} = \sum_{T_g, M_{T_g}'} A_{\nu E}^{T_g M_{T_g}'} (M_{T_e}) Y_{T_g M_{T_g}'}^*(\hat{k})$ we have \cite{pra:2012:deb}
  \begin{align}
     F^{T_g M_{T_g}'}_{T_g M_{T_g}}=\sum_{\substack{ \nu, M_{T_e}}} {A_{{\nu}E}^{T_g M_{T_g}'}\Lambda^{\nu, T_e M_{T_e}}_{T_g M_{T_g}}(E)}.
     \label{F}
     \end{align} 
Here  $\Lambda^{\nu, T_e M_{T_e}}_{T_g M_{T_g}}(E)$ is the amplitude for optical transition 
$\mid T_e,M_{T_e}\rangle _\nu \rightarrow 
\mid E, T_g M_{T_g} \rangle$ due to L$_{\nu}$ laser, where $\mid E, T_g M_{T_g} \rangle$  represents unperturbed (laser-free)
scattering state for $(T_g, M_{T_g})$ quantum numbers. Explicitly, 
 \begin{align} 
  A_{\nu E}^{T_g M_{T_g}'}(M_{T_e})=  D_{\nu}^{-1} \left [ \left \{ \Lambda^{\nu, T_e M_{T_e}}_{T_g M_{T_g}'}(E)\right \}^*
  + {\mathscr L}_{\nu}^{\nu'} \right ]\nonumber\\ \nu'\ne \nu
  \label{ane2}
 \end{align} 
 where $D_{\nu} = \zeta_\nu + i{\cal J}_\nu/2$  with ${\cal J}_\nu = \Gamma_\nu+\Gamma_{\nu\nu'}+\gamma_{\nu}$, 
 $\zeta_\nu = E + \hbar\Delta_\nu-(E_{\nu}+E_\nu^{{\rm shift}}+E_{\nu\nu'}^{{\rm shift}})$. Here 
 \begin{align} 
 {\mathscr L}_{\nu}^{\nu'} =  \xi_{\nu'}^{-1}{\cal K}_{\nu\nu'}.
\end{align} 
with 
$\xi_\nu (E) = E +  \hbar \Delta_{\nu} -  (E_{\nu}+E_{\nu}^{\rm{shift}}) + i \Gamma_{\nu}/2 $ 
and 
\begin{align} 
{\cal K}_{\nu \nu'}  = \left  ({\cal V}_{\nu \nu'} - i \frac{1}{2} \hbar {\cal G}_{\nu \nu'} \right )
\label{knnpll}
\end{align} 
is the cross coupling between two excited bound states
 induced by the two lasers, where 
\begin{align}
{\cal V}_{\nu \nu'} = \sum_{T M_{T}} {\mathcal P} \int \frac{ \Lambda_{ T_g M_{T_g}}^{\nu, T_e M_{T_e}}(E') 
\left \{\Lambda_{T_g M_{T_g}}^{\nu', T_{e} M_{T_e}}(E') \right \}^*} {E - E'} dE'\\
{\cal G}_{\nu \nu'} = \frac{2 \pi}{\hbar}  \sum_{T M_{T}} \Lambda_{T_g M_{T_g}}^{\nu, T_e M_{T_e}}(E) 
\left \{ \Lambda_{T_g M_{T_g}}^{\nu', T_{e} M_{T_e}}(E)\right \}^*.  
\end{align} 
The stimulated line width of $\nu$th bound state due to the $\nu$th laser only  is 
\begin{align}
\Gamma_\nu(E)& =& \frac{2\pi}{\hbar} \sum_{ M_{T_g}}
|\Lambda_{T _gM_{T_g}}^{\nu, T_e M_{T_e}}(E')|^2
\end{align} 
and the corresponding light  shift is 
 \begin{align}
 E_{\nu}^{{\rm shift}} = \sum_{ M_{T}} {\mathcal P} \int \frac{ \Lambda_{T_g M_{T_g}}^{\nu, T_e M_{T_e}}(E') 
\left \{ \Lambda_{T_g M_{T_g}}^{\nu, T_{e} M_{T{e}}}(E') \right \}^* } {E - E'} dE'.
  \end{align}
The terms that arise due to cross coupling ${\cal K}_{\nu \nu'}$ are  
\begin{align}
E_{\nu\nu'}^{{\rm shift}}= \text{Re}[\xi_{\nu'}^{-1}{\cal K}_{\nu\nu'}{\cal K}_{\nu'n}]\;\;\;\text{and}
\end{align} 
\begin{align} 
\Gamma_{\nu\nu'}=-2\text{Im}[\xi_{\nu'}^{-1}{\cal K}_{\nu\nu'}{\cal K}_{\nu'\nu}].
\end{align}

\section{Results and discussions}                                                    
  For numerical  illustration, we consider that the  two OFR lasers L$_1 $ and L$_2$ couple $T_g=2$ 
   to two PLR vibrational  states $\nu = 1$ ($b_1$) and $\nu=2$ ($b_2$),  respectively, with same 
 $T_e=3$. 
 As discussed in the previous section, 
 from symmetry considerations, the PLR states chosen will be accessible only from odd  partial-waves 
 (odd $\ell$) and nuclear spin triplet ($I=1$). As per the selection rules $\Delta T = 0, 1$ the excited rotational state $T_e = 3$
 will be accessible from the ground $T_g = 2$ or $T_g = 4$, which means from $\ell=1$ or $\ell=3$. We assume that 
 the contributions from $\ell=3$ is negligible due to low temperature. Further, we select laser detunings and intensities 
 such that the optical couplings to the levels $T_1 = T_2 = 1$ are negligible. We can thus restrict our study to $T_g=2$ only.  
   \begin{figure*}[ht]
	    \centering
        \subfloat{\includegraphics[width=0.45\textwidth]{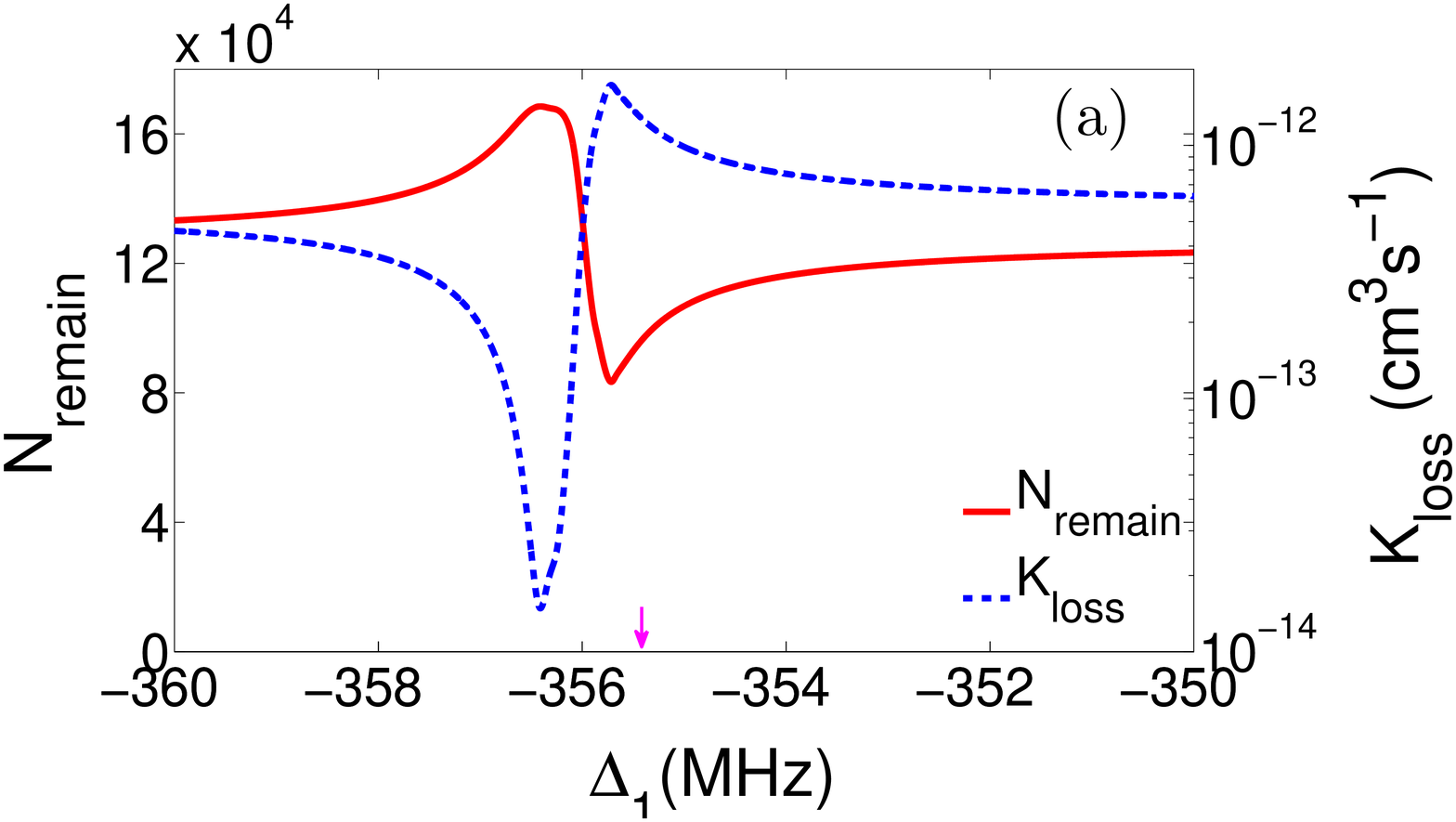}}\quad
        \subfloat{\includegraphics[width=0.45\textwidth]{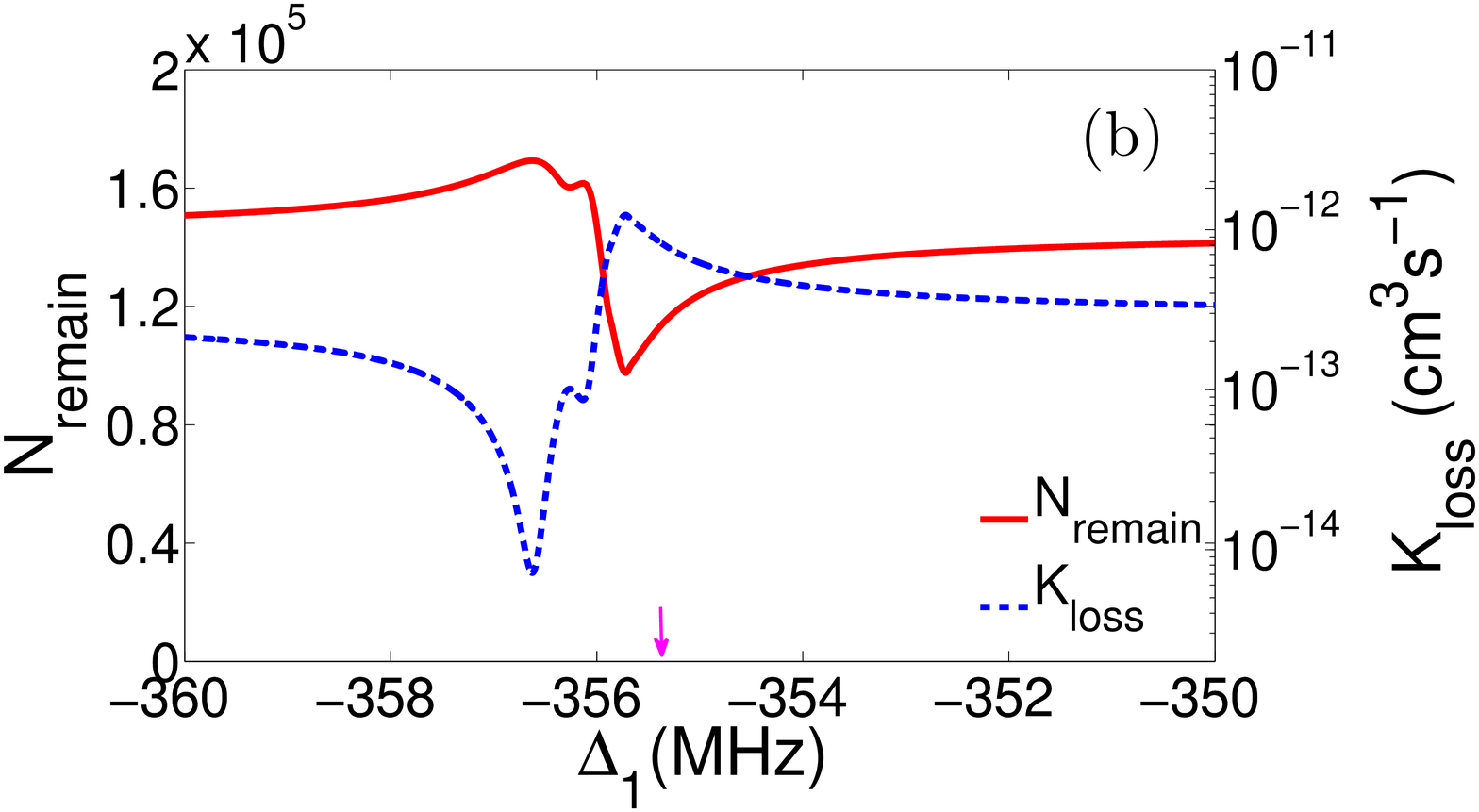}}\\
         \subfloat{\includegraphics[width=0.45\textwidth]{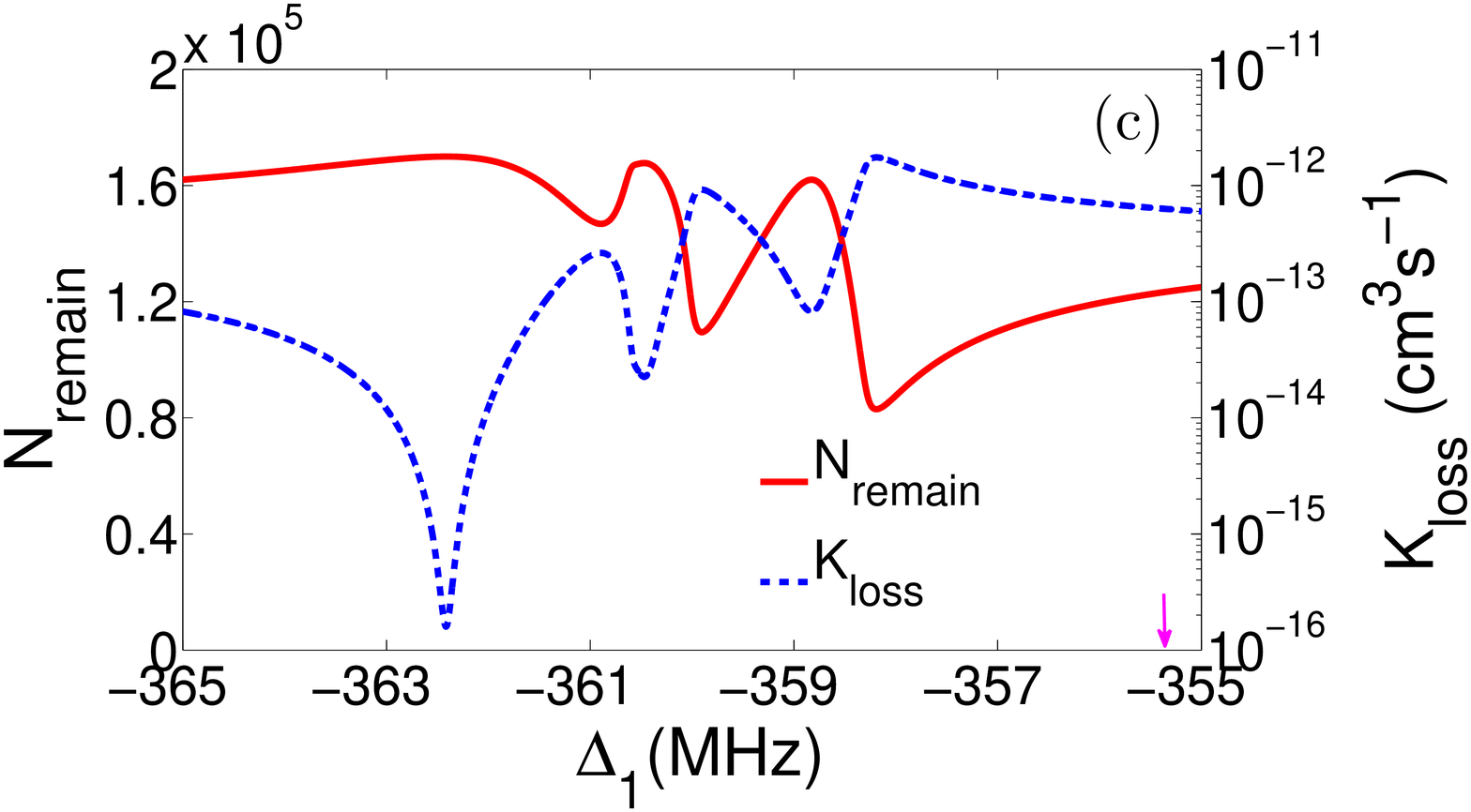}}\quad
        \subfloat{\includegraphics[width=0.45\textwidth]{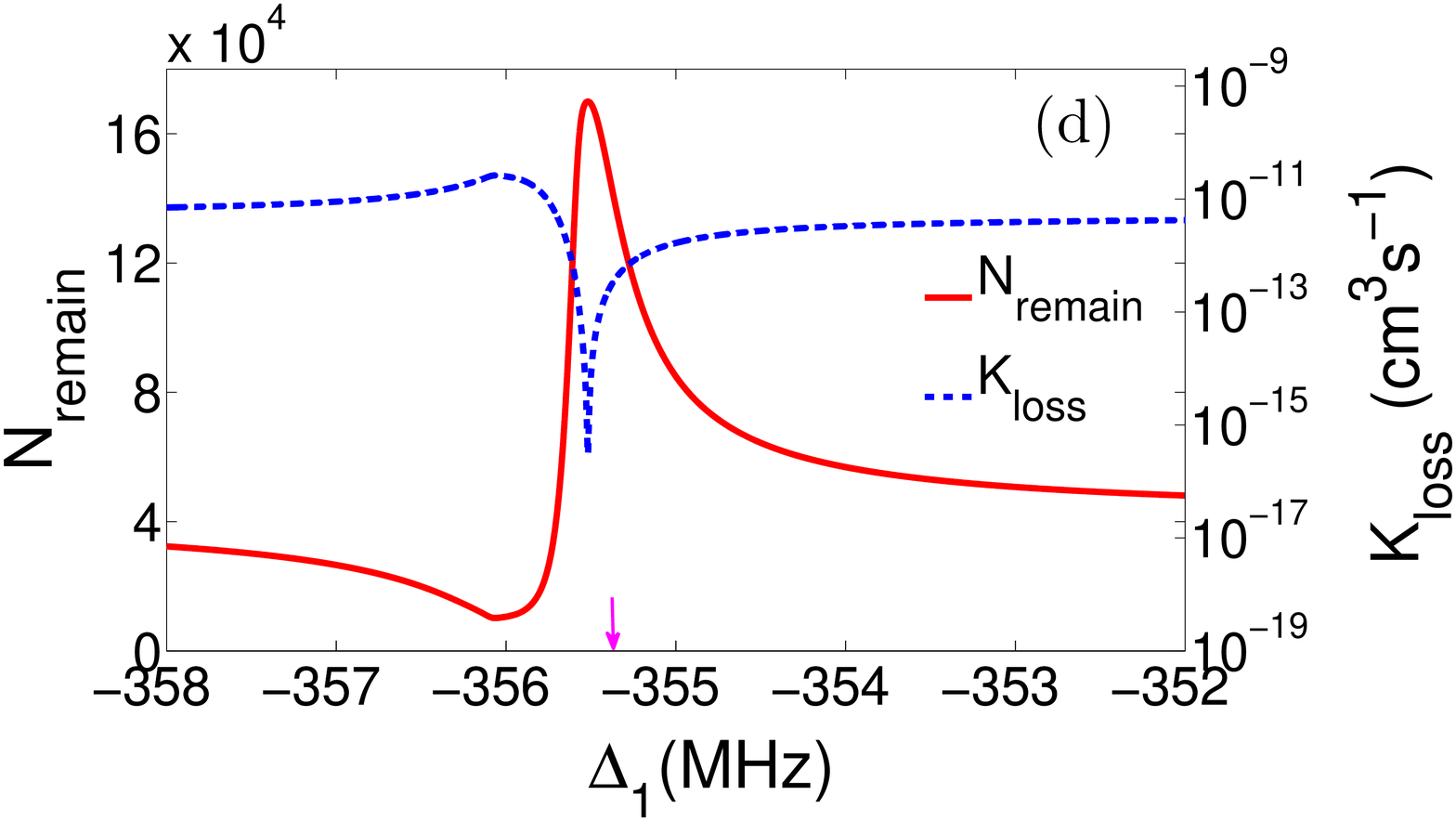}}

                     \caption{(Color online) Number of atoms  $N_{\rm {remain}} $ (solid) remaining in the trap after the two OFR lasers have acted for the time duration of 30 ms and PA rate $K_{\rm{loss}}$ (dashed) in unit of cm$^3$s$^{-1}$ are plotted against the detuning $\Delta_1$ (in MHz) of the first laser from the atomic transition frequency at temperature T = 8 $\mu$K for different laser intensities: (a) $I_1=1$Wcm$^{-2}$ ($\Gamma_1=212.9$ KHz)  and $I_2=10$Wcm$^{-2}$ ($\Gamma_2=4.3$ MHz) , (b) $I_1=1$Wcm$^{-2}$ ($\Gamma_1=212.9 $KHz )  and $I_2=20$Wcm$^{-2}$ ($\Gamma_2=8.6$ MHz ), (c) $I_1=10$Wcm$^{-2}$ ($\Gamma_1=2.1$ MHz ) and $I_2=20$Wcm$^{-2}$ ($\Gamma_2=8.6 $ MHz ), (d) $I_1=0.2$Wcm$^{-2}$ ($\Gamma_1=106$ KHz ) and $I_2=0.1$Wcm$^{-2}$ ($\Gamma_2=43 $ KHz ).The other parameters are: $\delta_2 = 0$ and laser phase difference $\theta=0$. The arrow indicates the binding energy $E_{b_1} = -355.4 $ MHz of the unperturbed bound state  $\mid b_1\rangle$.}                                       
                                              
      \end{figure*} 
  These two states have the binding energies  
 $E_{b_1}=-355.4$ MHz     $E_{b_2}=-212.4$ MHz   
 measured from the  threshold of excited potential.  
 The laser irradiation time is taken to be  30 ms, and the average atomic density $ \bar{n} =2\times10^{13}$ cm$^{-3}$  
 \cite{pra:2013:yamazaki,prl:2007:enomoto},  temperature is 8 $\mu$K and  
initial atom number $N_0=1.7\times10^5$.  Spontaneous line widths of the bound states are $\gamma_1 = \gamma_2 = 2\pi \times 364$ KHz \cite{pra:2013:yamazaki,prl:2007:enomoto}.

Fig. 3((a)-(d)) displays the number of atoms remaining $N_{{\rm remain}}$ and PA loss rate $K_{{\rm loss}}$ 
as a function of detuning of the first laser $\Delta_1 = \omega_{{\rm L}_1} - \omega_A $ 
from the atomic transition frequency $\omega_A$ (asymptote of the PLR potential) 
for different intensities of the two lasers. In Figs. 3(a) and 3(b), the first laser intensity is fixed in the intermediate coupling 
regime ($\Gamma_1 \le \gamma$) while the intensity of the second laser is set for 
the strong-coupling regime ($\Gamma_2 > \gamma$). For Fig. 3(c), the intensities of both lasers are in the strong-coupling 
regimes ($\Gamma_1 >\!> \gamma, \Gamma_2 >\!> \gamma$). $K_{{\rm loss}}$ (dashed curves) in Figs. 3(a)-(d) exhibit a 
prominent minimum, the values of  $K_{{\rm loss}}$  at the minimum  being about $3.8 \times 10^{-14}$ cm$^3$s$^{-1}$, 
$7.0 \times 10^{-15}$ cm$^3$s$^{-1}$  and $1.5 \times 10^{-16}$ cm$^3$s$^{-1}$, $2.96 \times 10^{-16}$ cm$^3$s$^{-1}$, respectively. 
Let $\Delta_{1min}$ be the value of $\Delta_1$ at which the minimum occurs.  
 $N_{{\rm remain}}$ as a function of $\Delta_1$ exhibits complementary behavior to that of $K_{{\rm loss}}$,  
 attaining the maximum exactly at  $\Delta_{1min}$. 
 The value 
$N_{{\rm remain}}$ at $\Delta_{1min}$  is nearly equal to  $ N_0$ implying that the loss of atoms is negligible at  $\Delta_{1min}$. 
This occurs due to the formation of molecular dark state 
leading  to the destructive quantum interference between  spontaneous emission transition 
pathways. The fact that the spectra for both $K_{{\rm loss}}$ and $N_{{\rm remain}}$ in 
Fig.3 are asymmetric with a prominent minimum  and a maximum is  indicative of the occurrence of quantum interference.
In case of one-laser OFR in the weak coupling 
regime ($\Gamma <\!< \gamma$), the spectra have symmetric Lorentzian shape. 
We notice that 
$\Delta_{1min}$ exhibits shifts towards the lower values of $\omega_{{\rm L}_1}$ as we move from Fig. 3(a) to Fig. 3(c). 
The increasing red shifts  due to increasing laser intensity $I_1$  are 
consistent with the calculated light shifts. Light shift $E_{\nu_n,M_{T_n}}^{\rm {shift}}$ of $\mid b_n\rangle$ due to L$_n$ laser only is propotional to the laser intensity $I_n$, and does not depend on the detunings. In contrast $ E_{\nu \nu'}^{\rm {shift}}$ depends on both laser intensities and the detuning $\Delta{\nu'}$. We have found $E^{{\rm shift}}_{\nu_1, M_{T_1} =0 }/I_1 =-1.62 $ MHzW$^{-1}$cm$^{2}$, $E^{{\rm shift}}_{\nu_1, M_{T_1} =1 }/I_1 =-1.44 $ MHzW$^{-1}$cm$^{2}$, $E^{{\rm shift}}_{\nu_1, M_{T_1} =2}/I_1 =-0.904 $ MHzW$^{-1}$cm$^{2}$  and $E^{{\rm shift}}_{\nu_1, M_{T_2} =0 }/I_2 = -1.82 $MHzW$^{-1}$cm$^{2}$, $E^{{\rm shift}}_{\nu_1, M_{T_2} =1 }/I_2 =-1.61 $ MHzW$^{-1}$cm$^{2}$, $E^{{\rm shift}}_{\nu_1, M_{T_2} =2}/I_2 =-1.00$ MHzW$^{-1}$cm$^{2}$. The calculated values of $ E_{12}^{\rm {shift}}$ and $ E_{21}^{\rm {shift}}$ are given in the table \ref{tab1} and \ref{tab2}, respectively.                  

   \begin{table}[h!]
        \begin{center}
           \caption{The values of  $ E_{12}^{\rm {shift}} (M_{T})$ in MHz are shown for  the laser intensities as used in Fig. 3 with $\delta_2=0$.}
          \vspace{4mm}
           \label{tab1}
       \begin{tabular*}{68mm}{| c|c | c|c | c|}
           \hline
        \hspace{2mm}$I_1$      &  \hspace{2mm}$I_2$ &\hspace{2mm} $ E_{12}^{\rm {shift}}$  \hspace{2mm}  & \hspace{2mm} $ E_{12}^{\rm {shift}}$     & $ E_{12}^{\rm {shift}}$   \\
         {W/cm$^2$}  & {W/cm$^2$} &($M_T=0$) &($M_T=1)$&( $M_T=2)$ \\    
         \hline
          1& 10& 1.09&0.975 &0.60\\
                            \hline
         1&20 &1.10&0.980&0.610\\
             \hline
            10& 20&11.03 &9.8 &6.10\\                    
            \hline
                         
            0.2& 0.1&0.117&0.098 &0.047\\                    
             \hline
           \end{tabular*}   
                     \end{center}
                     \end{table}

 \begin{table}[h!]
        \begin{center}
           \caption{The values of  $ E_{21}^{\rm {shift}} (M_{T})$ in MHz are shown for $\Delta_1=\Delta_{1min}$.}
          \vspace{4mm}
           \label{tab2}
       \begin{tabular*}{68mm}{| c|c | c|c | c|}
           \hline
        \hspace{2mm}$I_1$      &  \hspace{2mm}$I_2$ &\hspace{2mm} $ E_{21}^{\rm {shift}}$  \hspace{2mm}  & \hspace{2mm} $ E_{21}^{\rm {shift}}$     & $ E_{21}^{\rm {shift}}$   \\
         {W/cm$^2$}  & {W/cm$^2$} &($M_T=0$) &($M_T=1)$&( $M_T=2)$ \\    
         \hline
          1& 10&12.58&38.56 &-47.7\\
                            \hline
         1&20 &25.66&82.43&-77.72\\
             \hline
            10& 20&43.49 &41.52 &55.90\\                    
            \hline
                         
            0.2& 0.1&0.102&0.089&0.050\\                    
             \hline
           \end{tabular*}   
                     \end{center}
                     \end{table}         

From Fig. 3((a)-(c)) , we further notice that 
the width of the  dip in $K_{{\rm loss}}$ or equivalently the width of the maximum in $N_{{\rm remain}}$ 
increases as we move from Fig. 3(a) to Fig. 3(c). This means that the strong-coupling regimes with a molecular dark resonance 
are robust for efficient manipulation of atom-atom interactions. The oscillations in Fig. 3 result from the 
laser-induced coherence between the two excited bound states described by the term  ${\mathscr L}_{\nu}^{ \nu'}$. This laser 
induced coherence is important in the strong-coupling regimes because 
it comes into play due to nonlinear effects. One photon from L$_1$ laser excites the bound state $\mid b_1 \rangle$ which then  emits 
another photon by stimulated emission. When this absorption-emission cycle is followed by the excitation of the other 
bound state $\mid b_2 \rangle$ by one photon from L$_2$ laser, we have the coherence ${\mathscr L}_{\nu=2}^{ \nu'=1}$.
     \begin{figure}[hb]
	     \centering
               \includegraphics[width=0.48\textwidth]{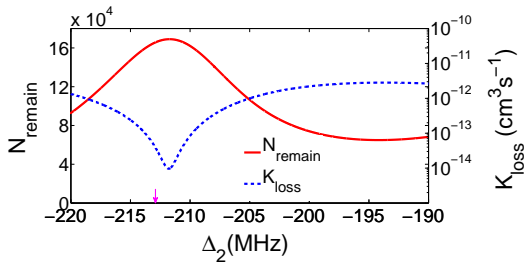}
               \caption{(Color online) Same as in  FIG. 3(b) but
               as a function of $\Delta_2$ at $\Delta_1=\Delta_{1min} $.}   
                \label{fig:n4}
                \end{figure}
The maximum in $N_{{\rm remain}}$ vs. $\Delta_1$ and the spectral asymmetry  
also appear in weak-coupling regimes for both lasers as shown in Fig. 3(d). At collision energy $E = 8 $ $\mu$K the square of the Franck-Condon overlap 
integral for the transition to $\mid b_2 \rangle$ is about 2 times larger than that to $\mid b_1 \rangle$. Therefore, if we fix the intensity of L$_1$ 
laser at half the intensity of L$_2$ laser, we expect that the two free-bound transition amplitudes with $M_{T_g} = M_{T_e}$
will be almost same. Then, in the weak-coupling regimes,  we would expect the maximum in  $N_{{\rm remain}}$ to occur at or near the energy of the 
unperturbed bound state  $\mid b_1 \rangle$ if the maximum arises due to the  dark state. In fact, the solid  curve in Fig. 3(d) has the maximum near the binding energy of $\mid b_1 \rangle$. 

Keeping the value of $\Delta_1$ fixed at $\Delta_{1min}$ which happens to be $\tilde{\delta}_1 \simeq 0$, 
we plot $K_{{\rm loss}}$ and $N_{{\rm remain}}$ as a function 
of $\Delta_2$ in Fig. 4. for the parameters as in Fig. 3(b).  This shows that  $K_{{\rm loss}}$ has a broad minimum when $\Delta_2$ is tuned near the resonance to $\mid b_2\rangle$. 
The loss of atoms is almost nil for the parameters at which the minimum in $K_{{\rm loss}}$ occurs. 

We next show elastic scattering cross section $\sigma_{el}$ as a function of $\Delta_1$ and compare it with the inelastic one $\sigma_{inel}$ 
in Fig. 5. keeping 
other parameters fixed as in Fig. 3(b). $\sigma_{el}$ is 6 orders of magnitude larger at and near $\Delta_{1min}$ where the minimum of $\sigma_{inel}$ 
occurs due to the dark state. Note that the back-ground (in the absence of OFR) elastic cross section is negligible ($\sim 10^{-19}$ cm$^2$). 
We have found that  $\sigma_{el}$ and $\sigma_{inel}$ are of comparable magnitudes in the weak-coupling regimes. \\

\begin{figure}[ht]
       \centering
       \includegraphics[width=0.48\textwidth]{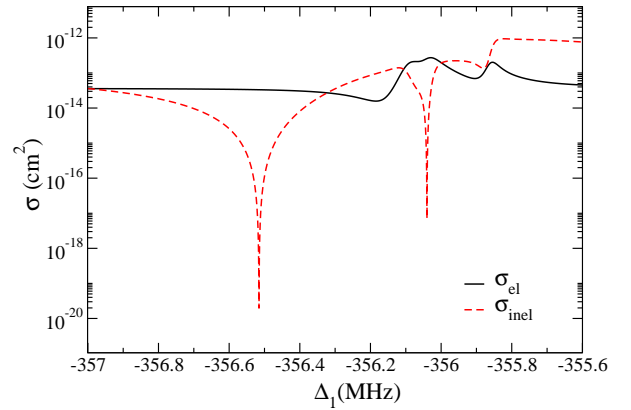}
        \caption{(Color online) $\sigma_{el}$ and $\sigma_{inel}$ are plotted as a function $\Delta_1$ for collision energy $E = 8 \mu$K.
      All other parameters are same as in Fig. 3(b). }
       \end{figure}

\section{conclusion}
In conclusion, we have demonstrated highly efficient manipulation of $p$-wave interactions between fermionic $^{171}$Yb atoms 
with a new optical method using two lasers in the strong-coupling regime. This method relies on creating a molecular dark state in 
the electronically excited potential, leading  to the  inhibition of photoassociative atom loss.
It is possible when two excited molecular bound states coupled 
to the continuum of scattering states by the two lasers have the same rotational quantum number;  and all pairs of excited sub-levels
having the same magnetic quantum number are  coupled a ground-state 
sub-level by the two lasers as schematically depicted in Fig. 2.
This ensures the cancellations of spontaneous emissions from all the excited sub-levels when 
the conditions for the formation of the dark state are fulfilled. The efficiency of our method may be characterized by a number of parameters:
(1) the ratio $N_{{\rm remain}}/N_0$ of the number of atoms remaining $N_{{\rm remain}}$ to the initial number $N_0$, 
(2) the ratio $\sigma_{el}/\sigma_{inel}$ of the elastic to inelastic 
scattering cross sections and (3) the ratio of the width of the dip in atom loss rate to the spontaneous line width. When the conditions 
for dark state are satisfied in the strong-coupling regimes for both the lasers, we have the results $N_{{\rm remain}}/N_0 \simeq 1$,  
$\sigma_{el}/\sigma_{inel} \sim 10^4 - 10^6$ and the ratio  between the two widths can be much larger than 1. Considering the velocity of the 
atoms to be a few cm s$^{-1}$, the number density $\sim 10^{13}$ cm$^{-3}$ we find elastic rate $\sim 1 $ s$^{-1}$ while the 
inelastic rate $\sim 10^{-5}$ s$^{-1}$. These numbers indicate that it is possible to manipulate atom-atom interactions 
 efficiently by the optical method presented in this paper. 
\section*{Acknowledgement}
 We are thankful to  Yoshiro Takahashi and Katsunari Enomoto for sending us a numerical code to calculate PLR potentials.

  \end{document}